\documentstyle[preprint,aps]{revtex}
\newcommand{\be}{\begin{equation}}
\newcommand{\ee}{\end{equation}}
\def\oA{{\bf A}}
\def\on{{\bf n}}
\def\oS{{\bf S}}
\def\oL{{\Lambda}}
\def\C{{\bf C}}
\def\oF{{\bf F}}
\def\R{{\bf R}}
\def\tA{{\tilde\oA}}
\def\tn{{\tilde\on}}
\def\tS{{\tilde\oS}}
\def\pdet{\hbox{``det''}}
\def\Tr{\hbox{\rm Tr}}

\bigskip

\tighten
\draft
\title{Coherent States of Alternating Current.\\ {\it (\today )}}
\author{ D. A. Ivanov$^{b}$, H. W. Lee$^a$, and L. S. Levitov$^{a,b}$}
\address{(a) Massachusetts Institute of Technology,
77, Massachusetts Avenue, Cambridge MA02139,\\
(b) L. D. Landau Institute for Theoretical Physics,
2, Kosygin str., Moscow 117334, Russia}

\begin{document}

\maketitle

\begin{abstract}
We study counting statistics of electric current pumped by
pulses of an external field. The fluctuations depend on the
pulse shape, and can be minimized by choosing the pulse shape
properly. For an optimal pulse shape, the fluctuations are
reduced to the {\it dc} level, i.e., they do not depend on the
duty cycle of the signal. We develop an approach that allows
to calculate all counting statistics for various driving fields,
optimal and non-optimal. The statistics depend in an interesting way on
the analytic structure of the field time dependence, and display
an analogy with coherent states and instantons.
  \end{abstract}
\narrowtext
\section{Introduction}

One of the fundamental problems of ``single electronics'' is that of
quantized charge transfer monitored by an external
field\cite{single_electronics}. A real device such as electronic
pump or turnstile is operated by a periodically alternating
field which drives electric current at a rate of one electron
per cycle\cite{turnstile}. Due to a microscopic scale of the
currently studied systems, the issue of current fluctuations,
classical and quantum, becomes very important.
Ideally, in order to reduce current
fluctuations well below the level of the {\it ac} signal,
the pump should transfer charge
adiabatically\cite{adiabatic_transmission}. For a perfectly
operating pump, the cycle duration $T$ is much longer than
$\hbar/E_c$, where $E_c$ is charging energy. If the adiabatic
limit $E_cT\gg\hbar$ is achieved, the
current fluctuations are exponentially
small.

In practice, the adiabatic parameter can be of the order of
one which makes the analysis very difficult, since one has to
consider many effects simultaneously:
co-tunneling\cite{co-tunneling},
finite relaxation rates\cite{Likharev},
quantum fluctuation of charge\cite{charge_fluctuation},
etc. In this paper we consider the problem in the non-adiabatic
limit $E_c\ll\hbar/T$, where the charging energy can be ignored.
With the usual assumption that Fermi-liquid quasiparticles'
interaction vanishes near Fermi surface, the pump can be
described by single particle scattering amplitudes periodically
varying in time. In this case the only effective mechanism of
the current fluctuation suppression is due to Fermi statistics
which makes electron transmission events correlated in the time
domain. The effect of Fermi statistics on the {\it dc} current
fluctuations has been studied\cite{quantum_noise}. We shall
extend this theory to the {\it ac} current.

{}From the point of view of the pump quality, the extreme
non-adiabatic limit we consider is the least efficiency mode.
So, it gives an upper estimate of the fluctuations at low
temperature, and provides a reference for understanding real systems
where non-adiabatic effects are strong. Actually, for the
non-interacting fermions one can develop a complete theory
which gives not just the mean square of current fluctuations,
but {\it all} statistics of transmitted charge\cite{IvanovL}.
Although the free fermion counting statistics problem captures
only part of the physics relevant for operation of a real
device, it is interesting enough and non-trivial by itself.

It turns out that the fluctuations strongly depend on the pulse
shape of the driving {\it ac} signal. The character of this
dependence resembles H-theorem. It was shown recently by two of us
that at fixed average current the fluctuations
level is bounded from below by the fluctuations of the {\it dc}
current  with  the  same   mean   value\cite{LeeLminim,LeeLortho}.
Moreover, there exist ``optimal'' {\it ac} signals for which the minimum
is reached, and the fluctuations remain on the {\it dc} level
(see Fig. 1). It is interesting that for optimal signals the
fluctuations are independent on the relative pulse width
$\tau/T$ (the signal duty cycle), no matter how sharp the
pulses are.

The case is appealing to an analogy with coherent states that
minimize quantum mechanical uncertainty. Similarly, the optimal
{\it ac} signals drive current in such a way that quantum noise
is reduced to the {\it dc} minimum level. Coherent states are
known to possess an interesting analytic
structure\cite{c-states}, and one encounters a similar situation
in this problem. We study analytic structure of the
problem, and show that it is related to the group of modular
transformations. The analytic structure facilitates the study
of counting statistics. We develop a method
discussed in our previous work\cite{IvanovL}, and use it to
calculate complete counting statistics for various interesting
examples of the driving signal.

To be specific, we study a one dimensional model of a Fermi gas
of electrons transferred through a potential barrier by an
external {\it ac} field. Electrons are incident from the left
and the right reservoirs which supply zero temperature Fermi
distribution. In the {\it ac} field, the scattering becomes
non-elastic, and electrons with different energies interfere due
to their Fermi-statistics. We treat the problem as a
multichannel scattering problem with the scattering amplitudes
given by the Fourier components of the external field, and
present a general formalism that gives counting statistics.

We compute the distribution exactly for a particular class of
external fields, which includes the optimal signals. For these
fields the law of time dependence is a periodic analytic
function of time given by a rational functions of the
``circular'' variable $z=e^{i\Omega t}$. We find exact
statistics of the charge transfer, and show that it displayes
interesting features explained by the analytic character of the
time-dependence law. In particular, we shall discuss Lorentzian
pulses of voltage with quantized flux: $c\int V(t)dt=n\Phi_0$,
where $n$ is an integer, and $\Phi_0={h c\over e}$ is the
flux quantum. We find that the pulses represent a quantum analog
of the classical picture of $n$ independent attempts to transmit
electrons through the scatterer. We arrive at this result by
means of the method of \cite{LesovikLbinom} combined with a
special treatment which allows one to reduce the problem in an
infinite-dimensional space to a finite dimensional problem (cf.
Ref.\cite{IvanovL}).

The paper is organized as follows. In sections 2 and 3 we
describe the model and state the mathematical problem to solve.
We introduce a generating function of the charge transfer
statistics and express it as the determinant of a scattering
operator in an infinite-dimensional space. For computing the
determinant it is important to find physically meaningful
regularization. The regularization problem is treated in section
4. We find that the regularization is sensitive to gauge
transformation of the electromagnetic field, and choose the
gauge so that the regularization becomes simple. In section 5 we
address the question of noise minimization and recall that the
signals that give minimal noise have particular analytic
structure: rational functions of the circular variable
$e^{i\Omega t}$ analytic inside the unit disc\cite{LeeLminim}.
We show how this result follows from our expression for the
generating function. Next, in section 6 we treat the case of a
rational time-dependence of the external field for which we
reduce the problem to finding the determinant of a
finite-dimensional matrix, and compute the probability
distribution explicitly. Section 7 contains the discussion of
the symmetries of the problem. We observe that the system
possesses a symmetry group of conformal transformations
isomorphic to $PSL(2,\R)$. Section 8 treats several interesting
examples of driving signal which reveal interesting features of
the counting statistics. In section 7 we summarize our
discussion. In order to make the readers' burden less heavy,
some technical details are moved from the main body of the paper
to the Appendices A and D. For completeness, we review relevant
results of \cite{LeeLminim} and \cite{LesovikLbinom} in the
Appendices B and C. In Appendix E we recall gauge
transformations that relate the {\it ac} flux problem to the
{\it ac} bias problem, which are essential for the issue of the
determinant regularization.

\section{Model}

We consider the following model of a microscopic contact.
One-dimensional ideal Fermi gas scatters off a potential barrier
$U(x)$, so that the Hamiltonian of the system is
  \begin{equation} H=\Psi^\dagger(x)\left[{1\over
2}\left(-i{\partial\over\partial x}
-{e\over c}a(x,t)\right)^2 +ev(x,t) +U(x)\right]\Psi(x),
\label{(1.a)}
\end{equation}
where $\Psi^\dagger(x)$ and $\Psi(x)$ are the canonical operators of
electrons, $U(x)$ is the scattering potential, $a(x,t)$ and $v(x,t)$ are
the external electromagnetic field vector and scalar potentials, respectively.
We shall treat
electrons as spinless and non-interacting. The scattering potential
$U(x)$ will be described by the matrix of scattering amplitudes
\begin{equation}  \oA=\pmatrix{A_{LL} & A_{RL} \cr A_{LR} & A_{RR}},
\label{(1.b)}
\end{equation}
so that the scattering states have the standard asymptotic form:
\begin{eqnarray}
\Psi_{L,k} &=\cases{e^{ikx}+&$ A_{LL}e^{-ikx},\quad x\to -\infty; $\cr
                            &$ A_{LR}e^{ikx},\;\quad x\to +\infty;$\cr}
\nonumber \\
\Psi_{R,k} &= \cases{ &$ A_{RL}e^{-ikx},\quad x\to -\infty; $\cr
            e^{-ikx} + &$ A_{RR}e^{ikx},\;\quad x\to +\infty.$\cr}
\label{(1.c)}
\end{eqnarray}

We would like to study the response of the system to a
time-dependent external field, electric or magnetic.
Let us recall that by a gauge transformation one can go from a
problem with the electric potential $v(x,t)$ to that with the vector potential
$a(x,t)$ and vice versa (see Appendix E). In the limit of {\it instant
scattering} which we shall assume from now on, the external field
may be taken into account by introducing a time-dependent phase
in the scattering amplitudes:
\begin{equation} \oA \to \tA(t)=
\pmatrix{A_{LL} & A_{RL}e^{-i\varphi(t)} \cr A_{LR}e^{i\varphi(t)} & A_{RR}}.
\label{(1.i)}
\end{equation}
``Instant scattering'' means that the external fields $a(x,t)$
and $v(x,t)$ (and thus $e^{i\varphi(t)}$ as well) vary
slowly in comparison with the scattering time
$\tau_{sc} \sim\hbar|\partial A_{\alpha\beta}/ \partial E|$, where
$A_{\alpha\beta}$ are the scattering amplitudes, $E$ is the energy of
electrons. The physical meaning of $\tau_{sc}$ is the time that the
particle spends inside the scatterer.
Technically, neglecting $\tau_{sc}$ means that the scattering amplitudes
$A_{\alpha\beta}$ do not depend on the energy.

\section{ Stating the problem}

Now, after the model is defined, we are ready to state the problem of the
counting
statistics of electron transmission. Namely, we are interested in
finding the probabilities of a given charge transfer over a fixed
time interval. We assume that the counting time $T^{(0)}$ is much longer than
the
period $T$ of the external field.
Let us observe the system under the action of the external
field
during a long time
interval
$T^{(0)}$.
The field from now on is treated only as the
phase factor $e^{i\varphi(t)}$ in Eq.(\ref{(1.i)}).
Let us denote the probabilities of transmitting exactly $n$
electrons
during this time by $P^{(0)}_n$. The probabilities can be conveniently
combined into a generating function
\begin{equation} \chi^{(0)}(\lambda)=\sum_{n=-\infty}^{+\infty} P^{(0)}_n
e^{i\lambda n}.
\label{(2.a)}
\end{equation}
This function of the auxiliary parameter $\lambda$ will be the main quantity
we shall work with. It encodes all the information about the statistics
of the charge transfer. The moments of the distribution are given by the
coefficients of the Taylor expansion of $\chi^{(0)}(\lambda)$ around
$\lambda=0$. In particular, the average charge transfer is
\begin{equation}\langle n
\rangle=-i{\partial\over\partial\lambda}\Big|_{\lambda=0}
\chi^{(0)} (\lambda),
\label{(2.b)}
\end{equation}
and the dispersion
\begin{equation}\langle\!\langle n^2 \rangle\!\rangle = \langle n^2 \rangle
-\langle n \rangle^2 = -{\partial^2\over\partial\lambda^2}
\Big|_{\lambda=0} \ln\chi^{(0)}(\lambda).
\label{(2.c)}
\end{equation}

Notice that even if we apply no external field there exist
equilibrium fluctuations. At zero temperature $\langle\!\langle
n^2\rangle\!\rangle$ is growing with time as $\ln T^{(0)}$
\cite{zeroTfermions}.
The zero temperature equilibrium fluctuations can be neglected
if the external field $e^{i\varphi(t)}$ is periodic in time: in
this case the field-induced fluctuations grow linearly with
$T^{(0)}$ and dominate over the equilibrium noise\cite{finiteT}.
In this paper we disregard the equilibrium noise and study only
the effect of the periodic field, and only briefly consider the
non-periodic case and the equilibrium noise in Appendix D.

There exists a useful description of the field-induced
statistics in terms of the quantities that do not depend on
the counting time. Let us introduce
  \begin{equation} P_n=\{\hbox{probability of transmitting $n$ electrons per
cycle}\}.
\label{(2.z)}
\end{equation}
The advantage of $P_n$ is that
the total generating function $\chi^{(0)}(\lambda)$ for a long time
interval $T^{(0)}\gg T$ can be written as an exponent:
\begin{equation} \chi^{(0)}(\lambda)\sim [\chi(\lambda)]^{T^{(0)}/T}
\qquad \hbox{as}\quad T^{(0)}\to\infty,
\label{(2.d)}
\end{equation}
where
\begin{equation}\chi(\lambda)=\sum_{n=-\infty}^{+\infty} P_n e^{i\lambda n}
\label{(2.e)}
\end{equation}
is the generating function per one cycle of the field.

This factorization can be explained by recalling that the zero
temperature equilibrium fluctuations are essentially due to the counting
beginning and ending. It is completely analogous to
the fluctuations in one-dimensional ideal Fermi gas (here the
role of the dimension is taken by the time). One can regard the
field-induced fluctuations as ``extensive in the time domain'',
and the equilibrium fluctuations as ``boundary effects''. As
usual, to treat the ``bulk'' effects separately from ``surface''
ones, we adopt periodic boundary conditions, which leads to
Eq.(\ref{(2.d)}).

More formally, let us first suppose that the phase factor is
periodic: $e^{i\varphi(t+T)}=e^{i\varphi(t)}$. Then the factorization
(\ref{(2.d)}) can be understood by the following argument. We
observe the system during a large number $N$ of field cycles and
impose boundary conditions periodic in time. The spectrum of
energies then is discrete with the spacing $\hbar\Omega/N=h/NT$.
Scattering is possible only between energy levels separated by a
multiple of $\hbar\Omega=h/T$. Thus, we have $N$ non-interfering
copies of the scattering problem with discrete energy levels
spaced by $\hbar\Omega$. Since $N=T^{(0)}/T$, this proves our formula
(\ref{(2.d)}).

If the phase factor $e^{i\varphi(t)}$ is not periodic, the
factorization still holds. In this case, since the voltage
${\hbar\over e}\dot\varphi$ is periodic, the phase factor is
quasi-periodic. The generating function of the statistics per
cycle is constructed in the following way. One has to find the
smallest positive $\epsilon$ for which the phase factor
$e^{i\varphi(t)+i\epsilon t}$ is periodic. Then $\chi(\lambda)=
\chi_1^{1-\epsilon}(\lambda) \chi_2^{\epsilon}(\lambda)$, where
$\chi_1$ and $\chi_2$ are found for the periodic phase factors
$e^{i\varphi(t)+i\epsilon t}$ and
$e^{i\varphi(t)+i(\epsilon-2\pi)t}$. For the proof and discussion
of the quasi-periodic case we refer to our previous
paper\cite{IvanovL}. In this paper only periodic phase
factors will appear.

The above argument shows that the field-driven contribution
$\chi(\lambda)$ can be extracted by closing the time axis into a
circle of period $T$, thus quantizing the energy with the
quantum $\hbar\Omega=h/T$. The problem becomes a
multi-channel scattering problem, where the channels represent
the discrete energy levels. For such a problem $\chi(\lambda)$
can be quite generally expressed in terms of scattering
amplitudes\cite{LesovikLbinom}. In this method the function
$\chi(\lambda)$ depends on a vector argument $\lambda
=(\lambda_1,\dots,\lambda_N)$, $N$ being the number of channels.
Fourier transform of $\chi(\lambda)$ gives the probabilities of
scattering between channels. If $\oA$ is the matrix of the
multi-channel scattering amplitudes, $\on$ is the occupation
number operator (diagonal in the energy representation),
$\oL=Diag(e^{i\lambda_1},\dots,e^{i\lambda_N})$, then
  \begin{equation}\chi(\lambda)=\det (1+\on(\oS-1)),
\label{(2.f)}
\end{equation}
where
\begin{equation}\oS=\oA^\dagger\oL^\dagger\oA\oL.
\label{(2.g)}
\end{equation}
The proof of this formula is reviewed in Appendix A.

In our treatment of the periodic problem the operators $\on$, $\oA$,
$\oL$, etc. act in the linear space
\begin{equation} H=V\oplus V
\label{(2.h)}
\end{equation}
of left and right states with discrete energies. Formally,
$V$ is the $\C^\infty$ space of the states of the discrete spectrum, and by
Fourier transformation it can be treated as the space of periodic
functions of time, with the period $T$.
Two copies of $V$ correspond to the left and right channels.
Further the computations will be performed in the basis $(\alpha,m)$, where
$\alpha\in\{R,L\}$ specifies the side, $m\in{\bf Z}$ labels the Fourier
harmonics. To be specific about notation, we shall
sometimes write operators $\oF$ acting in $H$ in the form
\begin{equation}\oF=\pmatrix{F_{LL} & F_{RL} \cr F_{LR} & F_{RR} \cr},
\label{(2.i)}
 \end{equation}
where $F_{\alpha\beta}$ are operators acting in $V\cong \C^\infty $.

In the application to our problem, the scattering matrix
is time dependent:
\begin{equation}\tA=\pmatrix{B&A^*[f(z)]^*\cr A f(z)&-B^*},
\label{(2.j)}
\end{equation}
where $A$ and $B$ are the transmission and reflection amplitudes
for the potential barrier $U(x)$, $|A|^2+|B|^2=1$;
$z=\exp(2\pi it/T)$ is the variable on the unit circle;
the phase factor $f(z)=\exp(i\varphi(t))$.
We choose the formal variables $\lambda_i$  to
count only total charge transfer,
regardless of the energy of transmitted electrons:
\begin{equation} \lambda_{L,m}=0, \qquad \lambda_{R,m}=\lambda\ .
\label{(2.k)}
 \end{equation}
Then from Eq.(\ref{(2.g)}) we get
\begin{equation} \tS=\tA^\dagger\oL^\dagger\tA\oL=
\pmatrix{|A|^2(e^{-i\lambda}-1)+1 &
A^*B^*(e^{i\lambda}-1)[f(z)]^* \cr -AB(e^{-i\lambda}-1)f(z) &
|A|^2(e^{i\lambda}-1)+1 \cr},
\label{(2.l)}
 \end{equation}
where the functions $f(z)$ and $[f(z)]^*=1/f(z)$ should be understood as
operators acting in $V$ by multiplication:
  \begin{equation}\langle m| f(z) |n\rangle
=\oint{dz\over 2\pi i} {f(z)\over z^{m-n+1}}.
\label{(2.m)}
\end{equation}
We consider the system at zero temperature, thus
\begin{equation} \on |\alpha,m\rangle=\cases{|\alpha,m\rangle, &$m\leq 0$;\cr
0, &$m>0$.}
\label{(2.n)}
 \end{equation}
By that, formally, all operators in (\ref{(2.f)}) and (\ref{(2.g)}) are
specified, and we can proceed with the calculation.

\section{Regularization of the determinant}

Now, the problem is to find the determinant (\ref{(2.f)}) of an
infinite matrix. Because of the infinite dimensionality, this
determinant needs to be understood properly. Eventually, to
compute the determinant, we are going just to cut a finite
submatrix of $1+\on(\tS-1)$ at some positive and negative
energies way off the Fermi level, and to take its determinant.
However, one has to be careful and to make sure that the
infinite parts of the matrix thrown away do not matter.
Physically, the reason is that all incident states with very high
energies are empty, and remain such after scattering off an
alternating field. Similarly, all states with very low energies
are deep in the Fermi sea, and remain doubly occupied throughout
the scattering. Thus, in order that our regularization procedure
has no effect on the determinant, we have to assure that the
unitarity of scattering is preserved when a submatrix is cut.
Then the infinite submatrices that we remove will contain only
the contribution of ``Fermi vacuum'' states, and will not affect
the counting statistics, and thus the determinant.

More formally, we have an infinite number of states below the
Fermi level in the left and the right reservoirs to be put in a
one to one correspondence. Shifting states in one reservoir with
respect to the states in the other one can be achieved by a
gauge transformations different for the left and the right
reservoirs. The key observation is that the matrix
$1+\on(\tS-1)$ is not invariant under a gauge transformation.
(The transformation rule of the particle number operator $\on$
is reviewed in Appendix E, and the matrices $\oA$ and $\tS$
transform the same way as the density matrix.) Of
course, since the determinant of $1+\on(\tS-1)$ gives counting
statistics, the determinant regularization must depend on the
gauge transformation in such a way that the regularized
determinant is gauge invariant.

In order to clarify the relation between the regularization and
the gauge transformations, let us consider a system with no
barrier. Then $A=1$, $B=0$, and the scattering is only forward,
no backward. Then the matrix $\tS$ given by (\ref{(2.l)})
becomes diagonal and time-independent:
  \begin{equation}
\tS= \pmatrix{e^{-i\lambda} & 0\cr 0 & e^{i\lambda}\cr},
\label{diagexample}
 \end{equation}
In this case, one could try to compute the determinant of
$1+\on(\tS-1)$ by using ``naive'' regularization, i.e., by
simply cutting all columns and rows above some large positive
and below some large negative energy. Then
$\chi(\lambda)={\rm det}(\tS)=1$: no transport for any
$\varphi(t)$, which is a meaningless result. The problem
becomes even more striking if one thinks of a gauge transformation.
Under a gauge transformation the $\tS$ given by (\ref{diagexample})
does not change, and $\on$ is transformed according to the rule
(\ref{(F.i)}) (see Appendix E). For example, if the gauge phase
$\phi(t)=n\Omega t$, in the energy representation one gets
\begin{equation}
\on'_L(E)=\on_L(E)\ ,\qquad \on'_R(E)=\on_R(E-n\hbar\Omega)\ .
\end{equation}
Now, by using the naive regularization one finds
\begin{equation}
\chi(\lambda)=e^{in\lambda}\ ,
\label{ambiguity}
\end{equation}
which means that the result is not gauge invariant.

At this point one can conclude that the correct regularization must change
together with the gauge. For the general scattering problem (\ref{(2.j)})
we proceed in the following way. We choose a gauge transformation
so that the scattering matrix $\oA$ becomes time
independent. Then $\tS$ is also time independent, and all phase
factors are shifted to $\on$. We argue that after such gauge
transformation one can use the naive regularization.
The transformation is chosen so that all dependence on the {\it ac} field is
moved
to the occupation of the incoming one-, two-, or many-particle
states. The advantage of going to a purely elastic scattering is
that all outgoing channels that enter the scattering unitarity
relation will have equal energies. Thus, while removing from
$1+\on(\oS-1)$ all states with energies below some large
negative energy we preserve the unitarity of scattering. In other words,
the states that interfere at the scattering are either both included in the
truncated matrix, or both are removed. Therefore, for our gauge transformation
the naive regularization is meaningful. Moreover, it is clear that such
transformation is unique, unless $\tS$ is diagonal\cite{LRambiguity}.

On the basis of this consideration, we argue that the
determinant (\ref{(2.f)}) must be understood in the following
way:
  \begin{eqnarray}
\chi(\lambda)
& =\pdet (1+\on(\tS-1))=
\pdet (1+\on\pmatrix{[f(z)]^*&0\cr 0&1} (\oS-1)
\pmatrix{f(z)&0\cr 0&1})
\nonumber
\\
& =\det (1+ \pmatrix{f(z)&0\cr 0&1}\on
\pmatrix{[f(z)]^*&0\cr 0&1}(\oS-1)) =
\det(1+\tn(\oS-1)),
\label{(2.o)}
\end{eqnarray}
   where
\begin{equation}\oS=\pmatrix{|A|^2(e^{-i\lambda}-1)+1 &
A^*B^*(e^{i\lambda}-1) \cr -AB(e^{-i\lambda}-1) &
|A|^2(e^{i\lambda}-1)+1 \cr}
\label{(2.p)}
 \end{equation}
is the time-independent scattering matrix, and
\begin{equation}\tn=\pmatrix{f(z)&0\cr 0&1}\on
\pmatrix{[f(z)]^*&0\cr 0&1}
\label{(2.q)}
\end{equation}
is the time-dependent occupation number operator. In the last
line of (\ref{(2.o)}) we can assign an unambiguous meaning
to the determinant (cf. \cite{IvanovL}). Namely, we note that
$\oS$ is unitary and has unit determinant. Also, $\tn$
tends to 0 at high energies and to 1 at low energies. Therefore,
the matrix $(1+\tn(\oS-1))$ behaves at infinity like a
block-diagonal matrix, with $2\times2$ blocks each having unit
determinant. We define the determinant in Eq.(\ref{(2.o)}) by
cutting the matrix along one of these blocks at infinity. In
such a way the determinant is well defined and depends
essentially on the matrix entries around the Fermi level. Below
we shall be able to compute it explicitly for a special choice
of $f(z)$.

\section{Noise minimization}

It turns out that the analytic structure of $f(z)$
plays an important role in the current fluctuations. In particular,
the functions $f(z)$ which minimize the noise
$\langle\!\langle n^2 \rangle\!\rangle$ for a given average charge transfer
$\langle n \rangle$ belong to the class of rational functions.
The problem of noise minimization has been considered in detail
in \cite{LeeLminim}, and here we briefly review the result.
The rate of charge transfer $\langle n \rangle$ is simply proportional
to the phase gain per period:
\begin{equation}\langle n \rangle =|A|^2 {\Delta \varphi\over 2\pi},
\qquad \Delta\varphi=\hbox{arg} f(z)\vert_{t=0}^{t=T},
\label{(3.a)}
\end{equation}
while the noise $\langle\!\langle n^2 \rangle\!\rangle$ depends on the
whole function $f(z)$:
\begin{equation}\langle\!\langle n^2 \rangle\!\rangle =2|A|^2|B|^2
\oint\!\oint {dz_1\, dz_2\over (2\pi i)^2} {f(z_1) f^*(z_2) -1
\over (z_1-z_2)^2}.
\label{(3.b)}
\end{equation}
For completeness, both expressions are derived in Appendix B.

The variational problem arises of finding the function $f(z)$ which
defines a map of a fixed degree $N=\Delta\varphi/2\pi$ of the unit circle
into itself and minimizes the noise functional (\ref{(3.b)}). In Appendix C
we review the proof of Ref.\cite{LeeLminim} that optimal $f(z)$ is
analytic either inside or outside the unit circle $|z|=1$.
In other words, its Laurent expansion $f(z)=\sum_{n=-\infty}^{+\infty}
c_n z^n$ contains either only non-positive or only non-negative
powers. Such a function can be written as
   \begin{equation} f(z)=\prod_{i=1}^N {z-a_i \over 1-a_i^*z},
\label{(3.c)}
 \end{equation}
where either all $|a_i|>1$, or all $|a_i|<1$. The corresponding
time dependence of the phase is:
   \begin{equation} \varphi(t) = \sum\limits_{i=1}^N
\tan^{-1}\left({(1-|a_i|^2)\sin\Omega(t-t_i)\over
(1+|a_i|^2)\cos\Omega(t-t_i)-2|a_i|}\right)\ +\ \phi_0\ ,
\label{kinks}
 \end{equation}
where $t_i=\hbox{arg}\,a_i/\Omega$, $\phi_0=\sum_i\hbox{arg}\,a_i$.
Thus the optimal phase time
dependence is a sum of $N$ ``elementary excitations'', or
``kinks'', each corresponding to a $2\pi$ phase change of the
scattering amplitude per one cycle of the signal.

For any of such functions the mean square fluctuation of the
transmitted charge $\langle\!\langle (en)^2 \rangle\!\rangle$ is
equal to $e^2|AB|^2N$ per cycle. It is remarkable that the noise does not
depend on relative displacement of the kinks in the time domain,
as well as on their durations. The degeneracy is described by $2N-1$
real parameters.

It is interesting that for the time dependence (\ref{(3.c)}) all
the probabilities $P_n$ can be computed and admit a simple
interpretation of $N$ non-interfering attempts of electrons to
pass through the barrier (see Examples 1 and 2 below). In fact,
the class of functions to which our method applies is broader:
it includes all rational functions (\ref{(3.c)}) regardless of
the location of the poles. (The phase time dependence then has
the form (\ref{kinks}) with arbitrary signs of different terms.)
We show that for any such $f(z)$ the generating function
$\chi(\lambda)$ can be expressed as the determinant of a {\it
finite} matrix, and that it yields only a {\it finite} number of
non-zero probabilities $P_n$.

\section{Computation}

Let $f(z)$ be a rational function that maps the unit circle into itself:
$|f(z)|=1$ for $|z|=1$. Then it has the form (\ref{(3.c)}) with arbitrary
$|a_i|\ne 1$, not necessarily all
inside or all outside the unit circle. Let
\begin{eqnarray}
Q(z)&=\prod_{i=1}^N (z-a_i),
\nonumber\\
          P(z)&=\prod_{i=1}^N (1-a_i^* z),
\label{(4.a)}
\end{eqnarray}
so that $f(z)=Q(z)/P(z)$.
We shall also use the functions
\begin{eqnarray}
Q^{-1}(z)={1\over Q(z)}=\sum_{k\in\bf Z}c_k^{(Q)}z^k,
\nonumber\\
          P^{-1}(z)={1\over P(z)}=\sum_{k\in\bf Z}c_k^{(P)}z^k,
\label{(4.b)}
\end{eqnarray}
where the Laurent expansions are chosen to converge on the unit circle.
We shall treat the functions (\ref{(4.a)}) and (\ref{(4.b)}) as operators
acting in $V$ by
multiplication.

Now we use the following trick to compute the determinant.
\begin{eqnarray}
\chi(\lambda) =\det (1+\tn(\oS-1)) &=
\det (1+\pmatrix{Q/P&0\cr 0&1}\on
\pmatrix{P/Q&0\cr 0&1}(\oS-1))
\nonumber\\
&= \det (1+\pmatrix{P^{-1}&0\cr 0&Q^{-1}}\on
\pmatrix{P&0\cr 0&Q}(\oS-1)).
\label{(4.e)}
\end{eqnarray}
Here we performed the gauge transformation in both left and right
channels simultaneously. This does not change the determinant.

Simple computations show that
the  matrices  $P^{-1}\on P$  and  $Q^{-1}\on Q$ have the
following form:
\begin{eqnarray}
P^{-1}\on P=\pmatrix{
\offinterlineskip
       \strut  \ldots & & &\ldots&\ldots&\ldots&\vrule&  &   &   \cr
       \strut         &1& &   *  &  *   &   *  &\vrule&  &   &   \cr
       \strut         & &1&   *  &  *   &   *  &\vrule&  &   &   \cr
       \strut         & & &p_{11}&\ldots&p_{1N}&\vrule&  &   &   \cr
       \strut         & & &\ldots&\ldots&\ldots&\vrule&  &   &   \cr
       \strut         & & &p_{N1}&\ldots&p_{NN}&\vrule&  &   &   \cr
                    \noalign{\hrule}
       \strut         & & &   *  &  *   &   *  &\vrule&  &   &   \cr
       \strut         & & &   *  &  *   &   *  &\vrule&  &   &   \cr
       \strut         & & &\ldots&\ldots&\ldots&\vrule&  &   &   \cr
},\\
Q^{-1}\on Q=\pmatrix{ \offinterlineskip
       \strut  \ldots & & &\ldots&\ldots&\ldots&\vrule&  &   &   \cr
       \strut         &1& &   *  &  *   &   *  &\vrule&  &   &   \cr
       \strut         & &1&   *  &  *   &   *  &\vrule&  &   &   \cr
       \strut         & & &q_{11}&\ldots&q_{1N}&\vrule&  &   &   \cr
       \strut         & & &\ldots&\ldots&\ldots&\vrule&  &   &   \cr
       \strut         & & &q_{N1}&\ldots&q_{NN}&\vrule&  &   &   \cr
                    \noalign{\hrule}
       \strut         & & &   *  &  *   &   *  &\vrule&  &   &   \cr
       \strut         & & &   *  &  *   &   *  &\vrule&  &   &   \cr
       \strut         & & &\ldots&\ldots&\ldots&\vrule&  &   &   \cr
},
\label{(4.f)}
\end{eqnarray}
blank spaces stand for zeroes, crosses mark the
Fermi level, asterisks denote arbitrary entries not used in
calculations.
Therefore
\begin{equation}\chi(\lambda) =\det\pmatrix
{\ldots &     &     &\ldots&\ldots&\ldots&   &   &   \cr
        & S_0 &     &  *   &  *   &   *  &   &   &   \cr
        &     & S_0 &  *   &  *   &   *  &   &   &   \cr
        &     &     &X_{11}&\ldots&X_{1N}&   &   &   \cr
        &     &     &\ldots&\ldots&\ldots&   &   &   \cr
        &     &     &X_{N1}&\ldots&X_{NN}&   &   &   \cr
        &     &     &  *   &  *   &   *  & I &   &   \cr
        &     &     &  *   &  *   &   *  &   & I &   \cr
        &     &     &\ldots&\ldots&\ldots&   &   & \ldots\cr }=
\det\pmatrix{ X_{11}&\ldots&X_{1N}\cr
                             \ldots &\ldots&\ldots\cr
                              X_{N1}&\ldots&X_{NN}\cr}.
\label{(4.g)}
\end{equation}
where
\begin{equation} X_{ij}=(\oS-I)\pmatrix{p_{ij} & 0 \cr 0 & q_{ij} \cr} +
I\delta_{ij}
\label{(4.h)}
\end{equation}
are $2\times 2$ matrices.

The determinant (\ref{(4.g)}) is finite and determines $\chi(\lambda)$ as a
function of the parameters $\{a_i\}$, $A$, and $B$.
As a function of $e^{i\lambda}$ and $e^{-i\lambda}$, the determinant
is a finite degree polynomial, since all entries of $X_{ij}$ are such
(see Eq.(\ref{(4.e)})). As a result, there is only a {\it finite} number of
non-zero probabilities $P_k$ in the expansion (\ref{(2.e)}) for
$\chi(\lambda)$.

\section{The $PSL(2,\R)$ symmetry}

Before we turn to the discussion of examples, let us mention that the system
possesses an interesting symmetry group $PSL(2,\R)$ defined as the group of
real
unimodular matrices.

The symmetries are realized by the group $G$ of linear-fractional
transformations preserving the unit disc $|z|<1$ (this group is
3-dimensional and isomorphic to $PSL(2,\R)$). We claim that:

\hfill\parbox{0.9\textwidth}{
If two phase factors $f(z)$ and $\tilde f(z)$ are related by such a
transformation ($\tilde f(z)=f(g(z))$ for some $g\in G$), then
the generating function of the distribution $\chi(\lambda)$ is the
same for $f(z)$ and $\tilde f(z)$.
}

Indeed, let us notice that $\chi(\lambda)$
is expressed as the determinant (\ref{(2.f)}) of an operator in the space of
functions
on the circle $|z|=1$. Also, any $g \in G$ commutes with
the occupation
number operator $\on$ at zero temperature ($g$ does not mix positive
and negative Fourier harmonics). If we perform the conjugation
by $g$, the expression (\ref{(2.f)}) remains the same, except that
$f(z)$ gets replaced by $\tilde f(z)$. The determinant however does not
change under a conjugation. This proves our statement.

Let us note
that $f(z)$ is uniquely determined by its zeroes $a_i$ (or, equivalently,
by its poles $1/a_i^*$). The sets of the parameters $a_i$ for $f(z)$ and
$\tilde f(z)$ are also related by $g$: $a_i=g(\tilde a_i)$. Therefore,
as a function of $\{a_i\}$, $\chi(\lambda)$ is invariant under
the simultaneous mapping of all $a_i$ by $g \in G$.

\section{Examples}

Let us illustrate our discussion by actual
computing the probability distribution according to Eq.(\ref{(4.g)})
for several specially chosen functions (\ref{(3.c)}).

{\it Example 1.} The simplest case is $N=1$.
 Obviously the transformation $a_i \mapsto 1/a_i $ just
switches the direction of the charge transfer.
Therefore, without loss of generality
let $|a|>1$.   Then by using the expansions (\ref{(4.b)}) one gets
\begin{eqnarray}
&q_{11}=1,
\nonumber\\
&p_{11}=0,
\nonumber\\
&\chi(\lambda)=|A|^2e^{i\lambda}+|B|^2.
\label{(6.a)}
\end{eqnarray}
Thus we have exactly one attempt to pass the barrier per period with
the probabilities $|A|^2$ to pass and $|B|^2$ to rebound. Such a situation
already occurred in the problem with constant voltage \cite{LesovikLbinom}.
Constant voltage is a special case of this example
corresponding to $a = 0$ or
$a = \infty $ (i.e., $f = \exp(\pm i\Omega t)$ ).

{\it Example 2.} Let now $N>1$, and the $a_i$, $i=1,...,N$, be all
inside or all
outside the unit disc (see Fig.
2.a). Again, without loss of generality let all $|a_i|>1$. Now,
we obtain
   \begin{eqnarray}
&q_{ij}=\delta_{ij},
\nonumber\\
&p_{ij}=0,
\nonumber\\
&\chi(\lambda)=(|A|^2e^{i\lambda}+|B|^2)^N.
\label{(6.b)}
\end{eqnarray}
We see that $\chi(\lambda)$ contains $N$ equal
factors, each
corresponding to one of the factors of
$f(z)$. This means that each ``elementary excitation'' in (\ref{kinks})
corresponds to one attempt to pass the barrier with the
one-particle outcome probabilities. Of course, the actual
scattering state in this case, as well as in Example 1,
is a many particle coherent state.
One could expect this result when all $a_i = \infty$ (this again
corresponds to constant voltage). However, here we get the same
expression for the superposition of {\it any} set of
``elementary excitations'' (Example 1) of the same polarization
with no dependence on the values of $a_i$. It is a surprising
result. Of course, in this example different sets of $\{a_i\}$
cannot be transformed into each other by the symmetry group $G$,
therefore we cannot explain this invariance merely by the
$PSL(2,\R)$ symmetry discussed previously. At present we look at
this invariance as at a miracle and admit this to be a
consequence of a broader group of symmetries.

We shall also list the answers for two other examples of $f(z)$
which demonstrate the interference of ``excitations'' with
opposite polarizations.

{\it Example 3.} $N=2$. $|a_1|>1$, $|a_2|<1$. Then one has
\begin{eqnarray}
 (p_{ij})={1\over a_1 ^*-a_2 ^*}\pmatrix{ a_1 ^* & -1 \cr
 a_1 ^*a_2 ^* & -a_2 ^* \cr},
 \nonumber\\
 (q_{ij})={1\over a_1 -a_2 }
 \pmatrix{ a_1  & -a_1 a_2  \cr 1 & -a_2  \cr},
 \nonumber\\
 \chi(\lambda)=1-2F+F(e^{i\lambda}+e^{-i\lambda}),
\label{(6.c)}
\end{eqnarray}
where
\begin{equation} F=|A|^2 |B|^2 {|1-a_1 ^*a_2 |^2 \over |a_1 -a_2 |^2}.
\label{(6.d)}
\end{equation}

Note that in this example the probabilities of the electron
transfer in both directions are equal, and thus average charge flux is zero,
although for arbitrary
parameters $a_1$ and $a_2$ the signal can be asymmetric (see Fig.
2.b). In this case $\chi(\lambda)$ cannot be represented as a
product of independent contributions of the two ``elementary
excitations'', but exhibit their interference. To make this
clear, let us show what happens if one tries to factor the
generating function:
  \begin{equation}
\chi(\lambda)= (u+w e^{i\lambda})(u+w e^{-i\lambda})\ ,
\label{notfactors}
\end{equation}
where the ``probabilities'' $u,w={1\over2}(1\pm\sqrt{1-4F}))$,
so in this case there is no natural relation between
the factors of $\chi(\lambda)$ and of $f(z)$.

{\it Example 4.} $N>1$. $a_1=a$,  $a_i=b$ for $i>1$.
$|a|>1$, $|b|<1$. Computations can be most easily performed for the case
$b=0$ (which corresponds to a constant voltage applied against one
``elementary excitation''). Then by using the $PSL(2,\R )$ symmetry
one can extend the result to arbitrary $b$:
\begin{equation} \chi(\lambda)=(1-2F+F(e^{i\lambda}+e^{-i\lambda}))
(|A|^2e^{i\lambda}+|B|^2)^{N-1},
\label{(6.e)}
 \end{equation}
where
\begin{equation} F=|A|^2 |B|^2 {|1-a^*b|^2 \over |a-b|^2}.
\label{(6.f)}
\end{equation}
  Like in Example 3, there is an interference of the kinks of
opposite sign. It is interesting that the factors of
$\chi(\lambda)$ can be interpreted by saying that one positive
and one negative kink interfere and form a ``neutral'' system
with which other kinks do not interfere. The degree of
interference is measured by $\Lambda=|1-a^*b|/|a-b|$, and varies
from 0 to 1 depending on how much the kinks overlap in time, or
by how much their durations differ. For example, $\Lambda\to1$
if the kinks almost not overlap, and then $\chi(\lambda)$
factors into separate contributions of independent kinks: the
``probabilities'' $u$ and $v$ in (\ref{notfactors})
become just the one particle
probabilities $|A|^2$ and $|B|^2$.

\section{Discussion}

To summarize, we found the probability distribution for the charge
transfer under the action of a periodic external field described
by a rational function of $z=e^{i\Omega t}$. According to our
previous remarks about the equivalence between magnetic and electric
fields (also, see Appendix E), we can treat the field $f(z)$
of the form (\ref{(3.c)}) as the alternating voltage $V(t)$:
\begin{eqnarray}
V(t)&=&-{i\over e}\, {d\over dt}\, \ln f(e^{i\Omega t})
={\Omega z\over e} \sum\limits_{i=1}^N {1-|a_i|^2\over 1+|a_i|^2-
(a_i^* z+ a_i z^{-1})}
\nonumber\\
&=&{\Omega\over e} \sum\limits_{i=1}^N {1-\rho_i^2\over 1+\rho_i^2
-2\rho_i\cos(\Omega t-\varphi_i)},
\label{(7.a)}
\end{eqnarray}
where $a_i=\rho_i e^{i\varphi_i}$.

This expression represents $V(t)$ as a sum of ``elementary
excitations''. Each elementary excitation alone represents
one attempt of an electron to pass the barrier,
with the probability to pass given by the transmission
coefficient of the barrier.
We found that if the excitations are of the
same sign, then these attempts do not interfere.
The general formula (\ref{(4.g)}) holds
for any $V(t)$ composed of elementary excitations of
arbitrary signs. It predicts a nontrivial interference of
the excitations of opposite signs.

Each elementary excitation in Eq.(\ref{(7.a)}) can be written as
a sum of Lorentzian pulses:
  \begin{equation}
V_k(t)=
{\Omega\over e} {1-\rho_k^2\over 1+\rho_k^2
-2\rho_k\cos(\Omega t-\varphi_k)}=\sum\limits_{m=-\infty}^{\infty}
{\Omega\over e}{2\tau_k\over (t-t_k-mT)^2+\tau_k^2}\ ,
\label{Lorentzian}
\end{equation}
with the width $\tau_k=\Omega^{-1}\ln(\rho_k^{-1})$, centered
at $t_k=\Omega^{-1}\varphi_k$. Each of the pulses carries flux
proportional to its area. The flux is quantized:
  \begin{equation}
c\int\limits_{-\infty}^{\infty}
{\Omega\over e}{2\tau_kdt\over (t-t_k-mT)^2+\tau_k^2}\ ={hc\over e}=\Phi_0\ .
\label{fluxquantum}
\end{equation}
The flux quantization is just another way to say that each
elementary excitation corresponds to a $2\pi$ phase shift.

Using the representation (\ref{Lorentzian})
our results can be to some extent translated to the non-periodic
case. The limiting form of the excitation as
the period $T\to\infty$ ($\Omega\to 0$) is
   \begin{equation} V(t)={2\hbar\over e}
{\tau\over (t-t_0)^2+\tau^2}\ ,
\label{(7.b)}
\end{equation}
a single Lorentzian pulse with the area $\int V(t)\, dt=h/e$. If many
such pulses all of the same sign are generated continuously at a finite rate
so that average current dominates over the equilibrium fluctuations,
then the calculation will tell that each pulse corresponds to a
``one-electron like'' attempt to pass the barrier, and the
distribution of outcomes is binomial (exactly as for constant
voltage). The interference of the field-driven current with the
equilibrium noise will be discussed elsewhere.


\section{Conclusion}

We  studied  quantum  counting statistics of an {\it ac} current
driven by pulses of external  field.  There  are  special  pulse
configurations  which create many particle coherent scattering
states in which the quantum  noise  is  reduced  to  a  {\it  dc}
minimum.  The  analytic structure of such states is studied, and
put in  connection  with  the  modular  symmetry  group  of  the
problem.  A  general  method to calculate counting statistics is
presented  and applied to the coherent states and to other states
``naturally related'' to them. The counting statistics are found
to be binomial and ``generalized binomial'', respectively.

\acknowledgements
Research of L.L. is partly supported by Alfred Sloan fellowship.
Research at the L. D. Landau Institute is supported by the
International Science Foundation grant \#M9M000.

\appendix
\section{Expression for the generating function}

In this Appendix we review the proof of (2.f) from \cite{LesovikLbinom}.
First we express the determinant (2.f) as
\begin{equation} \chi(\lambda)=\det(1-\on+\on\oS)=\sum_{\{i_1,\dots,i_k\}}
S^{i_1\dots i_k}_{i_1\dots i_k} \prod_{i\ne i_\alpha} (1-n_i)
\prod_{i=i_\alpha} n_i,
\label{(A.a)}
\end{equation}
where the summation is over all subsets of channels
$\{i_1,\dots,i_k\}$. Here $S^{j_1\dots j_k}_{i_1\dots i_k}$ denotes
the determinant of the submatrix of $\oS$ formed by the entries
in the rows $\{i_1,\dots,i_k\}$ and in the columns $\{j_1,\dots,j_k\}$.
{}From (\ref{(2.g)}) it follows that
\begin{equation} S^{i_1\dots i_k}_{i_1\dots i_k}=\sum_{\{j_1,\dots,j_k\}}
e^{i(\lambda_{j_1}+\dots+\lambda_{j_k}-\lambda_{i_1}-\dots-
\lambda_{i_k})}
|A^{j_1\dots j_k}_{i_1\dots i_k}|^2\ ,
\label{(A.b)}
\end{equation}
where the determinants
\begin{equation}
A^{j_1\dots j_k}_{i_1\dots
i_k}=\sum\limits_P\varepsilon(P)A_{i_1}^{j_{P(1)}}\dots
A_{i_k}^{j_{P(k)}}
\end{equation}
are $n-$particle scattering amplitudes. Note that
\begin{equation}P_{i_1\dots i_k|j_1\dots j_k} =
\prod_{i\ne i_\alpha}(1-n_i) \prod_{i=i_\alpha} n_i
|A^{j_1\dots j_k}_{i_1\dots i_k}|^2
\label{(A.c)}
\end{equation}
are the probabilities of the many-electron scattering from the
channels $i_1,\dots,i_k$ to the channels $j_1,\dots,j_k$. This
proves that
\begin{equation} \chi(\lambda)=\sum_{\{i_1,\dots,i_k\}\atop\{j_1,\dots,j_k\}}
P_{i_1\dots i_k|j_1\dots j_k}
e^{i(\lambda_{j_1}+\dots+\lambda_{j_k}-\lambda_{i_1}-\dots-
\lambda_{i_k})}
\label{(A.d)}
\end{equation}
is the generating function for the probability distribution
$P_{i_1\dots i_k|j_1\dots j_k}$ of the charge transfer.

\section{Expressions for $\langle n\rangle$ and
$\langle\!\langle n^2\rangle\!\rangle$}

{}From (\ref{(2.o)}) we can derive expressions for the total charge
transfer and its dispersion in terms of the external field
$e^{i\varphi(t)}$. We shall perform the calculations for the periodic
field $e^{i\varphi(t)}=f(e^{i\Omega t})$. One can write
\begin{eqnarray}
\langle n\rangle&=&-i{\partial\over\partial\lambda}\Big|_{\lambda=0}
\chi(\lambda)=-i{\partial\over\partial\lambda}\Big|_{\lambda=0}
\ln\chi(\lambda)=
-i\Tr {\partial\over\partial\lambda}\Big|_{\lambda=0}
(1+\tn(\oS-1))
\nonumber\\
&=&-i\Tr(\tn {\partial\over\partial\lambda}\Big|_{\lambda=0} (\oS-1))
=\Tr \left[\tn \pmatrix{-|A|^2 & -A^*B^*\cr -AB & |A|^2 \cr}\right]\ .
\label{(B.d)}
\end{eqnarray}
In the time representation
($z=e^{i\Omega t}$),
\begin{equation}
\on(z_1,z_2)=\sum\limits_{n\le0}z_1^{-n}z_2^ne^{\delta n}=
z_2{1\over z_2(1+\delta)-z_1}\ ,
\label{(B.e)}
\end{equation}
where $\delta$ is infinitesimal positive. Thus one gets
\begin{equation} \tn(z_1,z_2)=z_2 \pmatrix{{f(z_1)f^*(z_2)\over
z_2(1+\delta)-z_1}&0\cr 0&{1\over z_2(1+\delta)-z_1}\cr},
\label{(B.f)}
\end{equation}
Therefore,
\begin{eqnarray}
\langle n\rangle = |A|^2\oint\left(\lim_{z_2\to z_1} {f(z_1)
f^*(z_2) -1\over z_1-z_2}\right){dz_1\over 2\pi i}&
\nonumber\\
=-|A|^2\oint {dz\over 2\pi i} f(z)\partial f^*(z) =
|A|^2\oint {d\varphi(t)\over 2\pi} =|A|^2 {\Delta\varphi\over 2\pi},&
\label{(B.g)}
\end{eqnarray}
where $\Delta\varphi$ is the phase change per period.

Similarly,
\begin{eqnarray}
\langle\!\langle n^2\rangle\!\rangle &=&-{\partial^2\over\partial\lambda^2}
\Big|_{\lambda=0} \ln\chi(\lambda)=-\Tr
{\partial^2\over\partial\lambda^2} \Big|_{\lambda=0} \ln (1+(\tn
(\oS-1))
\nonumber \\
&=&-\Tr\left[\left(\tn{\partial\over\partial\lambda} \Big|_{\lambda=0}
(\oS-1)\right)^2-\tn
{\partial^2\over\partial\lambda^2} \Big|_{\lambda=0} (\oS-1)\right]
\nonumber \\
&=&\Tr\left[\tn \pmatrix{-|A|^2 & -A^*B^*\cr -AB & |A|^2 \cr}
\tn\pmatrix{-|A|^2 & -A^*B^*\cr -AB & |A|^2 \cr}
-\tn\pmatrix{|A|^2 & -A^*B^*\cr AB & |A|^2 \cr}\right]
\nonumber \\
&=&\oint{dz_1\over 2\pi i} \lim_{z_2\to z_1}\Bigl[
\oint {dz_3\over 2\pi i}\Bigl( |A|^4{f(z_1)f^*(z_2) +1\over
(z_1-z_3(1+\delta))(z_3-z_2(1+\delta))}
\nonumber \\
&\ &+ |A|^2 |B|^2{f(z_1)f^*(z_3)+f(z_3)f^*(z_2)
\over (z_1-z_3(1+\delta))(z_3-z_2(1+\delta))} \Bigr)
-{2|A|^2\over z_1-z_2(1+\delta)} \Bigr]
\nonumber \\
&=& 2|A|^2|B|^2\oint\!\oint {dz_1\, dz_2\over(2\pi i)^2}\,
{f(z_1)f^*(z_2)-1\over (z_1-z_2)^2}\ .
\label{(B.h)}
\end{eqnarray}
The last line shows that $\langle\!\langle n^2\rangle\!\rangle$
depends in a non-trivial way on the whole function $f(z)$,
unlike $\langle n\rangle$ which depends only on the total phase
shift $\Delta\varphi$ per period. This recovers the result of \cite{LeeLminim}
for a periodically varying field.

\section{Variational problem}

In this Appendix we review the proof of \cite{LeeLminim} that
the variational problem of minimizing $\langle\!\langle n^2
\rangle\!\rangle$ for a fixed value of $\langle n \rangle$ is
equivalent to the analyticity of $f(z)$ either inside or
outside the unit circle.

We decompose $f(z)$ into a sum of $f_+(z)$ and $f_-(z)$ which are
analytic inside and outside the unit circle, respectively,
\begin{eqnarray}
f_+(z)=\sum\limits_{n=0}^\infty a^+_n z^n,&
\nonumber\\
f_-(z)=\sum\limits_{n=0}^\infty a^-_n z^{-n}&.
\label{(C.a)}
\end{eqnarray}
Then, by Cauchy theorem,
\begin{equation}\oint{dz\over 2\pi i} f_+(z)\partial f^*_-(z)=
\oint{dz\over 2\pi i} f_-(z)\partial f^*_+(z)=0.
\label{(C.b)}
\end{equation}
Therefore,
\begin{eqnarray}
\langle n\rangle= -|A|^2
\oint{dz\over 2\pi i} f(z)\partial f^*(z)=
-|A|^2\oint{dz\over 2\pi i} [f_+(z)\partial f_+^*(z)
+f_-(z)\partial f_-^*(z)]&
\nonumber \\
= |A|^2\sum_n n (|a_n^+|^2-|a_n^-|^2)&,
\label{(C.c)}
\end{eqnarray}
and
\begin{eqnarray}
\langle\!\langle n^2 \rangle\!\rangle=
2|A|^2|B|^2\oint\!\oint {dz_1\, dz_2\over(2\pi i)^2}\,
{f(z_1)f^*(z_2)-1\over (z_1-z_2)^2}&
\nonumber \\
= |A|^2|B|^2\oint{dz\over 2\pi i}
\Bigl(\partial f_+(z) - \partial f_-(z)\Bigr) f^*(z)&
\nonumber \\
= |A|^2|B|^2\oint{dz\over 2\pi i}
\Bigl(\partial f_+(z) f^*_+(z) -
\partial f_-(z) f^*_-(z)\Bigr) &
\nonumber \\
= |A|^2|B|^2 \sum_n n (|a_n^+|^2+|a_n^-|^2)&.
\label{(C.d)}
\end{eqnarray}
By comparing the two expressions we see that at fixed
$\langle n \rangle$ the fluctuation
$\langle\!\langle n^2 \rangle\!\rangle$
is minimal when $f_+$ or $f_-$ vanishes: then
$\langle\!\langle n^2 \rangle\!\rangle_{min}=|B|^2|\langle n\rangle|$.

\section{Non-periodic signal}

Here we make a remark on how the formulas (\ref{(2.f)}), (\ref{(2.g)}) can
be applied to the case of a non-periodic field. Actually, the
expression (\ref{(2.f)})  for the multi-channel characteristic function
$\chi(\lambda)$
is quite general and can be applied to the non-periodic case as well.
Suppose the charge is measured during a finite time $T^{(0)}$. This
can be taken into account by making the parameter $\lambda$ of
the generating function time-dependent, and by ``turning it on''
only for the time interval of the measurement.
If the external field is encoded into a (non-periodic)
phase factor $f(t)$, then the distribution is given by expressions
(\ref{(2.o)}) and (\ref{(2.g)}), where
\begin{equation} \oL=\oL(t)=\pmatrix{e^{i\lambda(t)}&0\cr 0&1\cr},
\label{(D.a)}
\end{equation}
\begin{equation} \lambda(t)=\cases{\lambda,&$0<t<T^{(0)}$;\cr
                       0,&$t<0$ or $t>T^{(0)}$.}
\label{(D.b)}
\end{equation}
The operator $\oS(t)$ becomes
unity outside the observation time interval, and this makes
the determinant (\ref{(2.f)}) well defined. We postpone a
general discussion
of the non-periodic case to elsewhere. Here we only
demonstrate that this approach gives the correct answer for the
equilibrium noise $\langle\!\langle n^2 \rangle\!\rangle$ in the absence
of external field.

By the argument of Appendix B,
\begin{eqnarray}
\langle\!\langle n^2 \rangle\!\rangle&=&-\Tr
{\partial^2\over\partial\lambda^2}
\Big|_{\lambda=0} \ln(1+\tn(\oS-1))
\nonumber \\
&=&\int_0^{T^{(0)}} {dt_1\over 2\pi i}\, \lim_{t_2\to t_1}
\Bigl[ 2|A|^4 \int_0^{T^{(0)}} {dt_3\over 2\pi i}\, {1\over
(t_1-t_3+i0)(t_3-t_2+i0)}
\nonumber \\
&\ &-{2|A|^2\over t_1-t_2+i0} + 2|A|^2|B|^2
\int_0^{T^{(0)}} {dt_3\over 2\pi i}\, {1\over
(t_1-t_3+i0)(t_3-t_2+i0)} \Bigr]
\nonumber \\
&=&2|A|^2 \int_0^{T^{(0)}} {dt_1\over 2\pi i}\,
\lim_{t_2\to t_1} \Bigl[ \int_0^{T^{(0)}} {dt_3\over 2\pi i}\,
{1\over (t_1-t_3+i0)(t_3-t_2+i0)} - {1\over t_1-t_2+i0} \Bigr]
\nonumber \\
&=&-2|A|^2 \int_0^{T^{(0)}} {dt\over 2\pi i}\,
\int_{t<0 \atop t>T^{(0)}} {dt_3\over 2\pi i}\, {1\over
(t-t_3)^2} ={|A|^2\over 2\pi^2}\int_0^{T^{(0)}} dt\,
\Bigl[{1\over T^{(0)}-t}+{1\over t}\Bigr]
\nonumber \\
&=&{|A|^2\over\pi^2} \ln{T^{(0)}\over \tau_{sc}},
\label{(D.c)}
\end{eqnarray}
where $\tau_{sc}$ is the ultraviolet cutoff set by a
characteristic scattering time of the system $\hbar{\partial
A(E)\over\partial E}$.

It is straightforward to check that the fluctuations given by
the last line of Eq.(\ref{(D.c)}) agree with the Nyquist
equilibrium noise spectrum $S_{\omega}={e^2\over h} G(\omega)
|\omega|$. Indeed,
\begin{eqnarray}
\langle\!\langle n^2 \rangle\!\rangle&=&
\int_0^{T^{(0)}}dt \int_0^{T^{(0)}}dt'
\langle\!\langle j(t)j(t') \rangle\!\rangle
\nonumber\\
&=& \int {d\omega\over 2\pi}{|1-e^{i\omega t}|^2\over\omega^2}S_\omega
={|A|^2\over\pi^2} \ln{T^{(0)}\over \tau_{sc}}\ ,
\label{NyquistLog}
\end{eqnarray}
where $|A|^2=|A(E_F)|^2=G(\omega=0)$, and $\hbar/\tau_{sc}$ is
the ultraviolet frequency cutoff.

\section{Gauge transformations}

In this Appendix we show how by gauge transformations one can
switch between the problem with an electric potential and that
with a vector potential.
Recall that the gauge transformation
\begin{equation} \Psi=\tilde\Psi e^{-i\phi(t,x)}
\label{(F.a)}
\end{equation}
changes the vector potential $a(x,t)$ and the electric potential
$v(x,t)$ as
\begin{eqnarray}
\tilde a= a-{c\over e}{\partial\phi(t,x)\over\partial x},
\nonumber \\
\tilde v= v+{1\over e}{\partial\phi(t,x)\over\partial t}.
\label{(F.b)}
\end{eqnarray}
Assume that we apply a time-dependent magnetic field with the
vector potential $a(x,t)$ localized around the scatterer. For
example, this can be realized as a varying magnetic flux
threading a conducting loop with the contact. Then, by the gauge
transformation (\ref{(F.a)}) with
   \begin{equation} \phi(t,x)={e\over c}\int^x_{-\infty} a(t,x) \, dx
\label{(F.c)}
\end{equation}
we can turn to the problem with the zero vector potential and
the electric potential
\begin{equation} v(t,x)={1\over e}{\partial\phi(t,x)\over\partial t}.
\label{(F.d)}
\end{equation}
In this case, the gauge phase shift across the scatterer is
\begin{equation}\varphi(t)=\phi(t,x)|_{x=-\infty}^{x=+\infty}={e\over
c}\int^{+\infty}_{-\infty} a(t,x) \, dx,
\label{(F.e)}
\end{equation}
and it can be viewed as the time-dependent voltage
\begin{equation} V(t)={1\over e}{\partial\varphi(t)\over\partial t}
\label{(F.f)}
\end{equation}
applied to the contact.

Thus we recalled the familiar gauge transformation for the
one-particle problem. When looking at a many-body problem,
one also has to transform the density matrix and Green's functions according to
the rule (\ref{(F.a)}). Let us consider the
occupation number operator of a reservoir:
  \begin{equation} \hat\on(\omega)=\sum_k \delta(\omega-E_k)
  \hat\psi^\dagger_k \hat\psi_k,
\label{(F.g)}
\end{equation}
where the summation is over all reservoir states. In the determinant
we have $\on=\langle\hat\on\rangle$ averaged over actual distribution. In
the time representation
\begin{equation} \on(t,t^\prime)=\sum_k e^{i E_k(t-t^\prime)}
\langle \hat\psi^\dagger_k \hat\psi_k \rangle
=\sum\limits_k e^{iE_k(t-t')}n_F(E_k)\ ,
\label{(F.h)}
\end{equation}
where $n_F(E)$ is Fermi distribution
Under the gauge transformation (\ref{(F.a)}) with the gauge phase
(\ref{(F.c)}) the occupation number operator transforms as
\begin{eqnarray}
&\tn_L&(t,t^\prime)=\on_L(t,t^\prime)
\nonumber\\
&\tn_R&(t,t^\prime)=\on_R(t,t^\prime) e^{i[\varphi(t)-\varphi(t^\prime)]}
\label{(F.i)}
\end{eqnarray}
for the left and the right reservoirs respectively.

We assume that the scattering is instant, then the alternating
field effect on the scattering states is entirely determined by
the phase shift $\varphi(t)$. This means that the {\it ac}
magnetic field can be introduced simply by adding the phase in
the scattering amplitudes (\ref{(1.i)}) , while in the problem
with the electric field we have the energies of occupied states
in one reservoir to be shifted with respect to those in the
other one. By virtue of gauge invariance, these two formulations
are obviously equivalent, and we make use of it in the
discussion of the determinant regularization (see
Eq.({\ref{(2.o)})).

\vfil   \eject

\centerline {\bf Figure Captions}

\bigskip

{\noindent} {\it Figure 1 :} Voltage  time
dependence (\ref{(7.a)}),
(\ref{Lorentzian}), and
(\ref{kinks})
that  makes  current fluctuations minimal. Shown is the simplest
solution: periodic sequence of  Lorentzian  peaks  of  the  area
$h/e$  each. For the optimal pulse shape the fluctuations do not
depend on the pulse width $\tau$.

\bigskip

{\noindent} {\it Figure 2 :}
Voltage optimal time dependence corresponding to a pair of
$2\pi-$kinks of $\varphi(t)={e\over\hbar}\int_{-\infty}^t V(t')dt'$.
For the pulses of equal sign ({\it a}) the counting
statistics do not depend on the relative position of the pulses $t_{1,2}$
and on their duration $\tau_{1,2}$ (see Example 2,
Eq.(\ref{(6.b)})). For the pulses of opposite sign ({\it b})
average current is zero and other statistics show non-trivial
interference of the kinks
(see Example 3, Eqs.(\ref{(6.c)})), (\ref{(6.d)})).

\bigskip

\vfil   \eject


\begin{references}
   \bibitem{single_electronics}
{\it Single Charge Tunneling}, NATO ASI Series B{\bf 294},
eds. H. Grabert and M. H. Devoret
(Plenum, New York, 1992);
G. Sch\"on and A. D. Zaikin, Phys. Rep. {\bf 198}, p. 238 (1990)
   \bibitem{turnstile} L. J. Geerlings, V. F. Anderegg,
P. Holweg, J. E. Mooij, H. Pothier, D. Esteve,
and M. H. Devoret, Phys. Rev. Lett. {\bf 64}, p. 2691 (1990)
   \bibitem{adiabatic_transmission}
D. Esteve, in Ref.\cite{single_electronics}, p. 109
   \bibitem{co-tunneling} D. V. Averin and A. A. Odintsov,
Phys. Lett. A{\bf 140}, p. 251 (1989)
   \bibitem{Likharev} D. V. Averin and K. K. Likharev, in
{\it Mesoscopic Phenomena in Solids}, eds. B. L. Altshuler, P. A. Lee and
R. A. Webb, p. 167 (Elsevier, Amsterdam, 1991)
\bibitem{charge_fluctuation} K. A. Matveev, Sov. Phys. JETP {\bf 72},
p. 892 (1991)
   \bibitem{quantum_noise}
G. B. Lesovik, JETP Letters, {\bf 49}, p.594 (1989);
B. Yurke and G.P.Kochanski, Phys.Rev.{\bf 41}, p.8184 (1989);
M. B{\"u}ttiker, Phys.Rev.Lett., {\bf 65}, p.2901 (1990)
   \bibitem{LeeLortho}
H.W.Lee, L.S.Levitov, Orthogonality catastrophe in a  mesoscopic
conductor due to a time-dependent flux. preprint
   \bibitem{LeeLminim}
H.W.Lee, L.S.Levitov, Estimate of minimal noise in a quantum conductor,
preprint
\bibitem{c-states} coherent states
   \bibitem{IvanovL}
D.A.Ivanov,  L.S.Levitov,
JETP Letters {\bf 58}(6), p.461 (1993)
   \bibitem{LesovikLbinom}
L.S.Levitov and G.B.Lesovik,
Charge distribution in quantum shot noise,
JETP Letters {\bf 58} (3) pp.230-235 (1993)
   \bibitem{zeroTfermions}
The formfactor $ \langle\!\langle n_qn_{-q}\rangle\!\rangle$ of
the equilibrium density-density correlator in a Fermi gas at
$T=0$ is given by $|q|/2p_F$ at $|q|<2p_F$, $1$ at
$|q|>2p_F$. The logarithmic dependence of the fluctuation of the
number of particles in a finite interval is derived by looking
at a double integral of the formfactor, very similar to the
expression (\ref{NyquistLog}) for the Nyquist noise.
   \bibitem{finiteT}
At finite temperature, equilibrium fluctuations grow linearly in
time: $\langle\!\langle n^2\rangle\!\rangle\sim T^{(0)}$. However, for
a temperature much less than the mean voltage the effect on
the counting statistics will be small.
   \bibitem{LRambiguity}
Diagonal $\oS$ means that the scattering is purely forward, and
thus the left and the right states neither mix, nor interfere.
In this case there is an ambiguity in the regularization: the
truncation energies for the left and the right channels can be
chosen differently. However, the ambiguity is absent if there is
a backscattering, no matter how weak.

\end{references}
\end{document}